\input{aipcheck}

\documentclass[
    ,final            
  ]
  {aipproc}

\layoutstyle{6x9}

\begin{document}

\title{Relativistic Positioning Systems: current status
$\! \!$\footnote{This paper is about the future interest and present
state of some developments of the theory of relativistic positioning
systems. For this reason, its structure has much of a potpourri of
already published papers, all of them explicitly mentioned in the
References.}}


\author{Bartolom\'e Coll, Joan J. Ferrando and Juan A. Morales-Lladosa}
{address={Departament d'Astronomia i Astrof\'{\i}sica, \\Universitat de
Val\`encia, 46100 Burjassot, Val\`encia, Spain.\\
bartolome.coll@uv.es, joan.ferrando@uv.es, antonio.morales@uv.es}}

\begin{abstract}
A {\em relativistic positioning system} consists in a set of four
clocks broadcasting their respective proper time by means of light
signals. Among them, the more important ones are the {\em
auto-located positioning systems,} in which every clock broadcasts
not only its proper time but also the proper times that it receives
from the other three. At this level, no reference to any exterior
system (the Earth surface, for example) and no synchronization are
needed. The current status of the theory of relativistic positioning
systems is sketched.
\end{abstract}

\maketitle


\section{Introduction}

In astronomy, space physics and Earth sciences, the increase in the
precision of space and time localization of events is associated
with the increase of a better knowledge of the physics involved. Up
to now all time scales and reference systems, although incorporating
so-called `relativistic effects'  \footnote{There are not
`relativistic effects' in relativity, as they are not `Newtonian
effects' in Newtonian theory. There are defects in the Ptolemaic
theory of epicycles that could be corrected by Newtonian theory, and
there are defects in Newtonian theory that may be corrected by the
theory of relativity. But then their correct appellation is not that
of `relativistic effects' but rather that of `Newtonian defects'.
\label{fn:Nwdefects}} in their development, start from Newtonian
conceptions.

Nowadays, a fully relativistic approach to Global Navigation Systems
is becoming more and more urgent. A pioneering proposal to develop
such a relativistic theory was presented at the ERE--2000 (XXIII
Spanish Relativity Meeting celebrated in Valladolid)
\cite{Coll-ERE-2000}, and also in
\cite{Coll-Bruseles-2001,Coll-Bucarest-2002,coll-tarantola}. At the
year 2005, the International School on {\it Relativistic
Coordinates, Reference and Positioning Systems} took place at
Salamanca, where the status of the theory at that time and work in
progress were communicated and discussed
\cite{Escola,Tarantola-lliso}. Since then, progress on the subject
has been attained and published elsewhere
\cite{ToloERE05}-\cite{Minko}.

In Relativity, the space-time is modeled  by a four-dimensional
manifold. In this manifold, most of the coordinates are usually
chosen in order to simplify mathematical operations, but some of
them, in fact a little set, admit more or less simple physical
interpretations. What this means is that some of the ingredients of
such coordinate systems (some of their coordinate lines, surfaces or
hypersurfaces) may be {\em imagined} as covered by some particles,
clocks, rods or radiations submitted to particular motions. But the
number of such {\em physically interpretable} coordinate systems
that can be {\em physically constructed} in practice is strongly
limited.

In fact, among the at present physically interpretable coordinate
systems, the only one that may be generically constructed%
\footnote{Such a statement is not, of course, a {\em theorem},
because involving real objects, but rather an {\em epistemic}
assertion that results from the analysis of methods, techniques and
real and practical possibilities of physical construction of
coordinate systems at the present time. \label{fn:epistemic}}
is the one based in the Poincar\'e-Einstein protocol of
synchronization, also called {\em radar system}, which uses
two-way light signals from the observer to the events to be
coordinated \cite{Perlick}. Unfortunately, this protocol suffers
from the bad property of being a retarded protocol (see below).
Consequently, in order to increase our knowledge of the physics
involved in phenomena depending on the space-time localization of
their constituents and, in particular, in making relativistic
gravimetry, it is important to learn to construct physically {\em
good} coordinate systems of relativistic quality.

The study of space-time coordinate systems and the diverse protocols
associated with their physical construction is a broad and open
field in current physics \cite{Perlick}-\cite{Hehl}.
Here, we consider
those coordinate systems constructed from {\it relativistic
positioning systems}, that are basically defined by four clocks
(emitters) broadcasting their respective proper times by means of
electromagnetic signals. At each space-time event reached by the
signals, the received four times define the {\it emission
coordinates} of this event (with respect to the given positioning
system).%
\footnote{It would be to notice here that a similar construction may
be carried out with sub-luminous signals, both in Newtonian and
relativistic physics (see \cite{newtonia}). This idea allows us, for
example, to develop the Newtonian theory of ultrasonic positioning
systems and the corresponding relativistic one. \label{fn:NwRelEC}}

\section{Location Systems}

To clearly differentiate the coordinate system as a mathematical
object from its realization as a physical object,%
\footnote{Different physical protocols, involving different physical
fields or different methods to combine them, may be given for a
unique mathematical coordinate system.
\label{fn:different-protocols}}
it is convenient to characterize this physical object with a proper
name. For this reason, the physical object obtained by a peculiar
materialization of a coordinate system is called a {\em location
system} \cite{Coll-Bucarest-2002,coll-tarantola,ToloERE05}. A
location system is thus a precise protocol  on  a particular set of
physical fields allowing to materialize a coordinate system.

A location system may have some specific properties
\cite{Coll-Bucarest-2002, ToloERE05}. Among them, the more important
ones are those of being {\em generic}, i.e. that can be constructed
in any space-time of a given class, (gravity-){\em free}, i.e. that
the knowledge of the gravitational field is not necessary to
construct it,%
\footnote{As a physical object, a location system lives in the
physical space-time. In it, even if the metric is not known, such
objects as point-like test particles, light rays or signals follow
specific paths which, a priori, allow  constructing location
systems. \label{fn:location- lives}}
and {\em immediate}, i.e. that every event may know its coordinates
without delay. Thus, for example, location systems based in the
Poincar\'e-Einstein synchronization protocol (radar systems) are
generic and free, but {\em not} immediate.

Location systems are usually used either to allow a given observer
assigning  coordinates to particular events of his environment or to
allow every event of a given environment to know its proper
coordinates. Location systems constructed for the first of these two
functions, following their three-dimensional Newtonian analogues,
are called ({\em relativistic}) {\em reference systems}. In
relativity, where the velocity of transmission of information is
finite, they are necessarily not immediate. Poincar\'e-Einstein
location systems are reference systems in the present sense.

Location systems constructed for the second of these two functions
which, in addition, are generic, free and immediate, are called
\cite{Coll-Bucarest-2002}  ({\em relativistic}) {\em positioning
systems.} Since Poincar\'e-Einstein reference systems are the only
known location systems but they are not immediate, the first
question is if in relativity there exist positioning systems having
the three demanded properties of being generic, free and immediate.
The epistemic%
\footnote{The word {\em epistemic} is also used here in the sense of
the footnote \ref{fn:epistemic}. \label{fn:epistemic2}}
answer is that there exists a little number of them, their
paradigmatic representative being constituted by four clocks
broadcasting their proper times by means of electromagnetic signals.

In Newtonian physics, when the velocity of transmission of
information is supposed infinite, both functions, of reference and
positioning, are {\em exchangeable} in the sense that data obtained
from any of the two systems may be transformed in data for the other
one. But this is no longer possible in relativity, where the
immediate character of positioning systems and the intrinsically
retarded  character of reference systems imposes a strong decreasing
hierarchy. In fact, whereas it is impossible to construct a
positioning system starting from a reference system (by transmission
of its data), it is always possible and very easy to construct a
reference system from a positioning system (it is sufficient that
every event sends its coordinate data to the observer).

Consequently, in relativity the experimental or observational
context strongly conditions the function, conception and
construction of location systems. In addition, by their immediate
character, it results that whenever possible, there are positioning
systems, and not reference systems, which offer the most interest to
be constructed. For the Solar system, it has been recently proposed
a `galactic' positioning system, based on the signals of four
selected millisecond pulsars and a conventional origin
\cite{coll-tarantola}. For the neighborhood of the Earth, a primary,
auto-locating, fully relativistic, positioning system has also been
proposed, based on four-tuples of satellited clocks broadcasting
their proper time as well as the time they receive from the others.
The whole constellation of a global navigation satellite system
(GNSS), as union of four-tuples of neighboring satellites,
constitutes an atlas of local charts for the neighborhood of the
Earth, to which a global reference system directly related to the
conventional International Celestial Reference System (ICRS) may be
associated (SYPOR project; see
\cite{Coll-Bruseles-2001,Coll-Bucarest-2002, ToloERE05,Pascual}).

\section{Relativistic positioning systems}

What is the coordinate system physically realized by four clocks
broadcasting their proper time?  Every one of the four clocks
$\gamma_A$ ({\it emitters})  broadcasting his proper time $\tau^A,$
the future light cones of the points $\gamma_A(\tau^A)$ constitute
the coordinate hypersurfaces  $\tau^A=constant$ of the coordinate
system for some domain of the space-time. At every event of the
domain, four of these cones, broadcasting the times $\tau^A,$
intersect, endowing thus the event with the coordinate values
$\{\tau^A\}.$  In other words, the past light cone of every event
cuts the emitter world lines at $\gamma_A(\tau^A)$; then
$\{\tau^A\}$ are the {\it emission coordinates} of this event.

Let $\gamma$ be an observer equipped with a receiver allowing the
reception of the proper times $(\tau^A)$ at each point of his
trajectory. Then, this observer knows his trajectory in these
emission coordinates. We say then that this observer is a {\it user}
of the positioning system. It is worth pointing out that a user
could, eventually (but not necessary), carry a clock to measure his
proper time $\tau$.

A positioning system may be provided with the important quality of
being {\it auto-locating}. For this goal, the emitters have also to
be {\it transmitters} of the proper time they just receive from the
other three clocks, so that at every instant they must broadcast
their proper time {\em and} also these other received proper times.
Then, any user does not only receive the emitted times $\{\tau^A\}$
but also the twelve transmitted times $\{\tau^A_B\}$. These data
allow the user to know the trajectories of the emitter clocks in
emission coordinates.

The basic properties of the emission coordinates have been analyzed
in \cite{D4a} for the generic four-dimensional case. As already
mentioned, the coordinate system that a positioning system realizes
is constituted by four null (one-parametric family of)
hypersurfaces, so that its covariant natural frame is constituted by
four null 1-forms. Such rather unusual {\em real null frames} has
been considered in the literature but very sparingly.

Zeeman \cite{Zeeman} seems the first to have used, for a technical
proof, real null frames, and Derrick found them as a particular case
of {\em symmetric frames} \cite{Derrick}, later extensively studied
by Coll and Morales \cite{RS}. They were also those that proved that
real null frames constitute a causal class among the 199 possible
ones \cite{199}.  Coll \cite{c-luz} seems to have been the first to
propose the physical construction of coordinate systems by means of
light beams, obtaining real null frames as the natural frames of
such coordinate systems. Finkelstein and Gibbs \cite{Finkelstein}
proposed symmetric real null frames as a checkerboard lattice for a
quantum space-time. The physical construction of relativistic
coordinate systems `of GPS type', by means of broadcasted light
signals, with a real null coframe as their natural coframe, seems
also be first proposed by Coll
\cite{Coll-ERE-2000,Coll-Bruseles-2001,Coll-Bucarest-2002,ToloERE05}.
Bahder \cite{bahder} has obtained explicit calculations for the
vicinity of the Earth at first order in the Schwarzschild
space-time, and Rovelli \cite{Rovelli}, as representative of a
complete set of gauge invariant observables, has developed the case
where emitters define a symmetric frame in Minkowski space-time.
Blagojevi\`c et al. \cite{Hehl} analysed and developed the symmetric
frame considered in Finkelstein and Rovelli papers.

References \cite{Rovelli, Hehl, bahder}, as well as ours
\cite{coll-tarantola,ToloERE05,JoanERE05,Pozo-Warsaw05,D4a,D2a,D2b},
have in common the awareness about the need of physically
constructible coordinate systems (location systems) in experimental
projects concerning relativity. But their future role, as well as
their degree of importance with respect to the up to now usual ones,
depends on the authors. For example, Bahder considers them as a way
to transmit to any user its coordinates with respect to an exterior,
previously given, coordinate system. Nevertheless, our analysis,
sketched above, on the generic, free and immediate properties of the
relativistic positioning systems lead us to think that they are
these systems which are assigned to become the {\em primary} systems
of any precision cartography. Undoubtedly, there is still a lot of
work to be made before we be able, as users, to verify and to
control this primary character, but the present general state of the
theory and the explicit results already known in two, three and
four-dimensional space-times encourage this point of view.  A key
concept for this primary character of a system, although not
sufficient, is the already mentioned of auto-location, whose
importance in the two-dimensional context has been shown (see next
section).

At first glance, relativistic positioning systems are nothing but
the relativistic model of the classical GPS (Global Positioning
System) but, as explained for example in \cite{ToloERE05}, this is
not so. In particular, the GPS uses its emitters (satellites) as
simple (and `unfortunately moving'!) beacons to transmit {\em
another} spatial coordinate system (the World Geodetic System 84)
and an ad hoc time scale (the GPS time), different from the proper
time of the embarked clocks, meanwhile for relativistic positioning
systems the unsynchronized proper times of the embarked clocks
constitute the fundamental ingredients of the system. As sketched in
\cite{Coll-Bucarest-2002} or \cite{ToloERE05}, positioning systems
offer a new, paradigmatic, way of decoupling and making independent
the spatial segment of the GPS system from its Earth control
segment, allowing such a positioning system to be considered as {\em
the primary positioning reference} for the Earth and its
environment.

\section{Two-dimensional approach} \label{2L-approach}

A full development of the theory for the generic four-dimensional
relativistic positioning systems requires a hard task and a previous
training on simple and particular situations. In Coll {\it et al.}
\cite{D2a} (see also Ferrando \cite{JoanERE05,JoanERE07}) we have
presented a two-dimensional approach to relativistic positioning
systems introducing the basic features that define them. This
two-dimensional approach has the advantage of allowing the use of
precise and explicit diagrams which improve the qualitative
comprehension of general four-dimensional positioning systems.
Moreover, two-dimensional scenarios admit simple and explicit
analytic results.

In a two-dimensional space-time we have two emitters $\gamma_1$ and
$\gamma_2$. Suppose they broadcast their proper times $\tau^1$ and
$\tau^2$ by means of electromagnetic signals, and that the signals
from each one of the world lines reach the other. The future light
cones (here reduced to pairs of `light' lines) cut in the region
between both emitters and they are tangent outside. Then, this
internal region, bounded by the emitter world lines, defines the
{\em emission coordinate domain} $\Omega$. Indeed, the past light
cone of every event in $\Omega$ cuts the emitter world lines at
$\gamma_1(\tau^1)$ and $\gamma_2(\tau^2)$, respectively. Then
$(\tau^1,\tau^2)$ are the coordinates of the event (see Fig. 1a).

Emission coordinates are null coordinates and thus, in the
two-dimensional case, the space-time metric depends on the sole {\em
metric function} $m$ and takes the expression: $ds^2 =
m(\tau^1,\tau^2) d \tau^1 d\tau^2$. The plane
$\{\tau^1\}\times\{\tau^2\}$ in which the different data of the
positioning system can be transcribed is called the {\it grid} of
the positioning system.

Here, a {\em user} of the positioning system corresponds, according
to the above four dimensional definition, to an observer $\gamma,$
traveling throughout an emission coordinate domain $\Omega$ and
equipped with a receiver allowing the reading of the received proper
times $(\tau^1,\tau^2)$ at each point of his trajectory. Any user
receiving continuously the emitted times $\{\tau^1, \tau^2\}$ knows
his trajectory in the grid. Indeed, from these {\em user positioning
data} $\{\tau^1, \tau^2\}$ the equation $F$ of the user trajectory
may be extracted: $\tau^2 = F(\tau^1)$.
\begin{figure}
\includegraphics[width=0.68\textwidth]{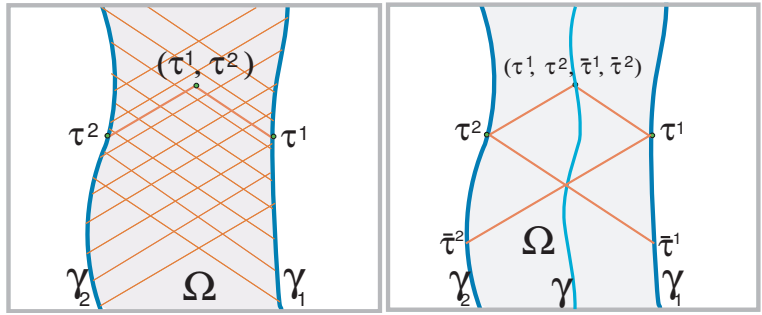}
%
\caption{\baselineskip 12pt (a) The region $\Omega$, bounded by the
emitter world lines, defines the emitter coordinate domain. (b) In
an auto-locating positioning system any user $\gamma$ receives from
the emitters the two proper times $\{\tau^1,\tau^2\}$ and the two
transmitted times $\{\bar{\tau}^2, \bar{\tau}^1\}$.}
\end{figure}
Analogously, {\em auto-locating positioning systems} are here
systems in which every emitter clock not only broadcasts its proper
time but also the proper time that it receives from the other, i.e.
the systems that broadcast the emission coordinates of the emitters.
Thus, the physical components of an auto-locating positioning system
are a {\em spatial segment} constituted by two emitters $\gamma_1$,
$\gamma_2$ broadcasting their proper times $\tau^1,$ $\tau^2$ {\em
and} the proper times $\bar{\tau}^2$, $\bar{\tau}^1$ that they
receive each one from the other, and a {\em user segment}
constituted by the set of all users traveling in an internal domain
$\Omega$ and receiving these four broadcast times $\{\tau^1, \tau^2;
\bar{\tau}^1, \bar{\tau}^2\}$ (see Fig. 1b). Then, any user
receiving continuously these {\em emitter positioning data} may also
extract from them the equations of the trajectories of the emitters
in emission coordinates: $ \varphi_1(\tau^1) = \bar{\tau}^2 \, , \
\, \varphi_2(\tau^2) = \bar{\tau}^1$.

Eventually, the positioning system can be endowed with complementary
devices. Thus, the emitters $\gamma_1$, $\gamma_2$ could carry
accelerometers and broadcast their acceleration $\alpha_1$,
$\alpha_2$, meanwhile the users $\gamma$ could be endowed with
receivers able to read, in addition to the emitter positioning data,
also the broadcast emitter accelerations $\{\alpha_1, \alpha_2\}$.
The users can also generate their own data, carrying a clock to
measure their proper time $\tau$ and/or an accelerometer to measure
their proper acceleration $\alpha$.%
\footnote{They can also carry gradiometers, but this situation will
be consider elsewhere.}

The purpose of the (relativistic) theory of positioning systems is
to develop the  techniques necessary to determine the space-time
metric as well as the dynamics of emitters and users by means of
physical information carried by the user data  $\{\tau^1, \tau^2;
\bar{\tau}^1, \bar{\tau}^2; \alpha_1, \alpha_2; \tau, \alpha\}$,
among others.

In the framework of this two-dimensional approach we have studied a
clarifying case: the positioning system defined by two inertial
emitters in flat space-time (Coll {\it et al.} \cite{D2a}). We have
obtained the coordinate change between emitter coordinates and
inertial null coordinates and, by using this change, we have shown
that the emitter trajectories in the grid are two straight lines
with complementary slope. Moreover, we have also found that the
emitter positioning data at two events determine the space-time
metric in emission coordinates (see Coll {\it et al.} \cite{D2a} for
more details).

\ \\
\noindent {\bf A. Gravimetry and positioning}  \ \\[2mm]
%
%
The interest of auto-locating positioning systems in gravimetry has
been pointed out in Coll {\it et al.} \cite{D2a}, where we have
shown that, if a user has neither a priori information on the
gravitational field nor on the positioning system, the user data
determine the metric and its first derivatives along the user and
emitter trajectories.

In a subsequent work (Coll {\it et al.} \cite{D2b}) we have gone
further in analyzing the possibility of making relativistic
gravimetry or, more generally, the possibility of obtaining the
dynamics of the emitters and/or of the user, as well as the
detection of the absence or presence of a gravitational field and
its measure. This possibility has been examined by means of a (non
geodesic) {\em stationary positioning system}, that is to say a
positioning system whose emitters are uniformly accelerated and the
radar distance from each one to the other is constant. Such a
stationary positioning system is constructed in two different
scenarios: Minkowski and Schwarzschild planes.

In both scenarios, and for any user, the trajectory of the emitters
in the {\em grid}, i.e. in the plane $\{\tau^1\}\times\{\tau^2\},$
are two parallel straight lines. Thus, we find that a stationary
positioning system may provide the same emitter positioning data
independently of the Schwarzschild mass. At first glance, this fact
would seem to indicate the impossibility of extracting dynamical or
gravimetric information from them, but this appearance is deceptive.

Indeed, we have proved in Coll {\it et al.} \cite{D2b} that the
simple qualitative information that the positioning system is
stationary (but with no knowledge of the acceleration and mutual
radar distance of every emitter) and that the space-time is created
by a given mass (but with no knowledge of the particular stationary
trajectories followed by the emitters) allows to know the actual
accelerations of the emitters, their mutual radar distances and the
space-time metric in the region between them in emission coordinates
$\{\tau^i\}.$ The important point for gravimetry is that, in the
above context, the data of the Schwarzschild mass may be substituted
by that of the acceleration of one of the emitters. Then, besides
all the above mentioned results, including the obtaining of the
acceleration of the other emitter, the actual Schwarzschild mass of
the corresponding space-time may be also calculated.

These relatively simple two-dimensional results strongly suggests
that relativistic positioning systems can be useful in gravimetry at
least when parameterized models for the gravitational field are
used.

\ \\
\noindent {\bf B. Positioning in flat space-time} \ \\[2mm]
%
%
Above we have considered stationary or geodesic positioning systems
where the user has, a priori, a partial or full information about
the positioning system. Now we consider a new situation: the user
knows the space-time where he is immersed, but he has no information
about the positioning system. Can the user data determine the
characteristics of the positioning system? Can the user obtain
information on his local units of time and distance and his
acceleration?

The answer to these questions is still an open problem for a generic
space-time, but we have solved it for Minkowski plane. In this flat
case we have analyzed the minimum set of data that determines all
the user and system information (see Coll {\it et al.} \cite{D2c}).
From the dynamical equation of the emitters and user we obtain that
the user data are not independent quantities: the accelerations of
the emitters and of the user along their trajectories are determined
by the sole knowledge of the emitter positioning data and of the
acceleration of only one of the emitters and only during an {\em
emitter echo}, i.e. the interval between the emission time of a
signal by an emitter and its reception time after being reflected by
the other emitter (see Fig. 2).

All the results obtained in this two-dimensional approach to
relativistic positioning systems show two things. Firstly, the
interest of this study in gravimetry and in building a fully
relativistic global navigation satellite system (GNSS). Secondly,
that a lot of work remains to be done in the development of the
general theory, the present one being only one of the first little
pieces. In the future, we want to put and solve new two-dimensional
problems that improve the qualitative comprehension of the
positioning systems, and we want to study the generalization of all
these two-dimensional results to the generic four-dimensional case.
\begin{figure}
\includegraphics[width=0.73\textwidth]{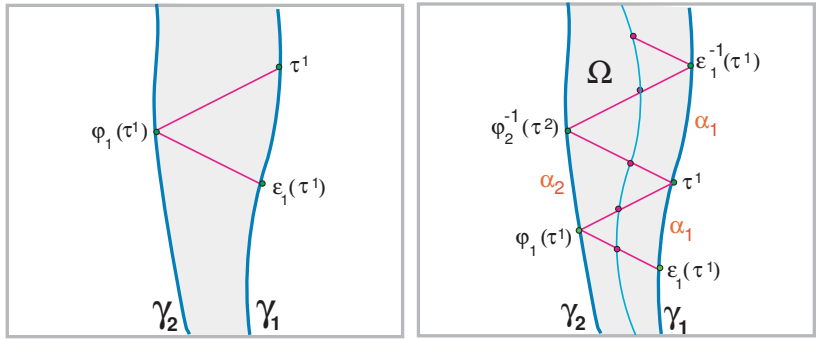}
\caption{\baselineskip 12pt (a) If an emitter $\gamma_1$ receives at
time $\tau^1$ a signal after being echoed by the other emitter
$\gamma_2$, it must be emitted at time $\epsilon_1(\tau^1)$,
$\epsilon_1 = \varphi_2 \circ \varphi_1$. (b) If a user receives an
emitter acceleration in the echo-causal interval
$[\epsilon_1(\tau^1), \tau^1]$, then he knows all the user data
along his trajectory.}
\end{figure}
\section{Emission coordinates in Minkowski space-time}

A user of a positioning system that receives the four times
$\{\tau^A\}$ knows its own coordinates in the emission system. Then,
if the user wants to know its position in another reference system
(for example the ICRS) it is necessary to obtain the relation
between both coordinate systems.

Thus, we have the following important problem in relativistic
positioning: suppose that the world-lines of the emitters
$\gamma_A(\tau^A)$ are known in a coordinates system
$\{x^{\alpha}\}$. Can the user obtain its coordinates in this system
if he receives its emission coordinates $\{\tau^A\}$? More
precisely, can the coordinate change $x^{\alpha}=x^{\alpha}(\tau^A)$
be obtained?

This coordinate change has been obtained for the case of emitters in
particular inertial motion in flat space-time \cite{BiniMas}.
Recently, we have solved this problem for a general configuration of
the emitters in Minkowski space-time \cite{NosERE08, Minko}. The
transformation $x^{\alpha}=x^{\alpha}(\tau^A)$ between inertial and
emitter coordinates is obtained in a covariant way depending on the
world-lines $\gamma_A(\tau^A)$ of the emitters.

In our study, the causal character (time-like, null or space-like)
of the emitter configuration plays an important role. The geometric
interpretation of this fact is a work in progress. In the short we
also want: (i) to study the expression of the coordinate change in a
3+1 formalism adapted to an arbitrary inertial observer, (ii) to
particularize our results for emitter motions modeling a satellite
constellation around the Earth, and (iii) to analyze the effect of a
week gravitational field on the coordinate transformation.

This current work is necessary to tackle in the future the stated
problem for a realistic situation modeling a constellation of a
GNSS.

\small{  \baselineskip 13pt \ \\{\bf Acknowledgements:} This work
has been supported by the Spanish Ministerio de Educaci\'on y
Ciencia, MEC-FEDER project FIS2006-06062.
}

\end{document}